\documentclass[conference]{IEEEtran}
\IEEEoverridecommandlockouts
\usepackage{cite}
\usepackage{url}
\usepackage{balance}
\usepackage{amsmath,amssymb,amsfonts}
\usepackage{algorithmic}
\usepackage{graphicx}
\usepackage{textcomp}
\usepackage{xcolor}
\usepackage{color}
\usepackage{listings}
\usepackage{tikz}
\usepackage{pgfplots}
\usepackage{booktabs}
\usepackage{url}
\usepackage[colorlinks = true,
            linkcolor = blue,
            urlcolor  = blue,
            citecolor = blue,
            anchorcolor = blue]{hyperref}

\def\BibTeX{{\rm B\kern-.05em{\sc i\kern-.025em b}\kern-.08em
    T\kern-.1667em\lower.7ex\hbox{E}\kern-.125emX}}

\definecolor{backcolour}{rgb}{0.95,0.95,0.92}

\lstdefinestyle{mystyle}{
    backgroundcolor=\color{backcolour},   
    basicstyle=\footnotesize,
    breakatwhitespace=false,         
    breaklines=true,                 
    captionpos=b,                    
    keepspaces=true,                 
    numbersep=5pt,                  
    showspaces=false,                
    showstringspaces=false,
    showtabs=false,                  
    tabsize=2
}

\newcommand{\newcomer}{newcomer candidate}
\newcommand{\Newcomer}{Newcomer candidate}

\definecolor{codegreen}{rgb}{0,0.6,0}
\definecolor{codegray}{rgb}{0.5,0.5,0.5}
\definecolor{codepurple}{rgb}{0.58,0,0.82}
\definecolor{backcolour}{rgb}{0.95,0.95,0.92}

\graphicspath{{pics/}}

\begin{document}

\title{Newcomer Candidate: Characterizing Contributions of a Novice Developer to GitHub}


\author{\IEEEauthorblockN{Ifraz Rehman, 
Dong Wang,
Raula Gaikovina
Kula, Takashi Ishio
and Kenichi Matsumoto
}
\IEEEauthorblockA{\textit{Nara Institute of Science and Technology, Nara, Japan}\\
\textit{Email: \{rehman.ifraz.qy4, wang.dong.vt8, raula-k, ishio, matumoto\}@is.naist.jp}}
}

\maketitle

\begin{abstract}
\textit{Context:} To attract, onboard, and retain any newcomer in Open Source Software (OSS) projects is vital to their livelihood.
Recent studies conclude that OSS projects risk failure due to abandonment and poor participation of newcomers.
Evidence suggests more new users are joining GitHub, however, the extent to which they contribute to OSS projects is unknown.

\textit{Objective:} In this study, we coin the term \textit{`\newcomer'}~to describe new users to the GitHub platform.
Our objective is to track and characterize their initial contributions.
As a preliminary survey, we collected 208 newcomer candidate contributions in GitHub.
Using this dataset, we then plan to track their contributions to reveal insights.

\textit{Method:} We will use a mixed-methods approach, i.e., quantitative and qualitative, to identify whether or not newcomer candidates practice social coding, the kinds of their contributions, projects they target, and the proportion that they eventually onboard to an OSS project.

\textit{Limitations:} The key limitation is that our newcomer candidates are restricted to those that were collected from our preliminary survey.
\end{abstract}

\section{Introduction}
\label{sec:introduction}

The success of Open Source Software (OSS) has always been the continuous influx of newcomers and their active involvement~\cite{park2009VISSOFT}.
Recent studies show evidence that many contemporary projects risk failing, with one of the core reasons is the inability to attract and retain newcomers~\cite{Fang2009JMIS,Steinmacher2014,Valiev2018FSE}. 
For example, Coelho and Valente~\cite{Coelho2017FSE} proposed two strategies that include newcomers, which aim {to transfer the project to new maintainers} and to {accept new core developers}.
In another study, Steinmacher et al.~\cite{Steinmacher2014} presented a model that analyzed the forces that are influential to newcomers being drawn or pushed away from a project.

The term \textit{newcomer} has usually been used in a loose way in literature \cite{igor2014IST}.
Newcomers can either be novice developers who are starting their career, experienced developers from industry that are new to OSS projects, or developers migrating from other OSS projects.
Most work are revolved around when newcomers have been onboarded into the OSS project.
Inspired by incubation projects of OSS, we coin the term:
\begin{quote}
   \textit{``A \textbf{\newcomer} is a novice developer that is a new user to the GitHub platform, with the intention to later onboard an OSS project''}
\end{quote}
Interestingly, GitHub reported 10 million new users in 2019\footnote{Statistics from \url{https://octoverse.github.com} accessed January 2020}.
With this upsurge in \newcomer~activity, the nature of their contributions is unknown, especially the extent to which their contributions assist OSS livelihood and sustainability. Furthermore, as a social coding platform, GitHub\footnote{\url{https://github.com}} allows developers to showcase their skills to the world's largest community of developers, attracting over 44 million repositories and engaging over 40 million plus developers.
Although there is a complete body of work that have studied the barriers and struggles of newcomers, none has looked at all contribution activities to both project and non-project repositories. 

To fill this gap between \newcomer{}s and their effects on OSS sustainability, in this registered report, we lay the foundation by tracking and characterizing \newcomer~contributions on GitHub.
As a preliminary survey, we have already collected and verified 208 \newcomer~contributions.
Using this dataset, we then plan to mine and track contributions to every repository.
Our protocol plan includes a mixed-methods approach with four research questions.
First, we identify {to what extent} do \newcomer{}s practice social coding with other users.
We then manually characterize the kinds of initial contributions.
Next, we characterize the kinds of projects that attract the interest of a \newcomer. 
Finally, we identify to what extent does a  \newcomer~eventually gets onboarded to an OSS project.
We envision that the results of our study will reveal insights on what contributions attract \newcomer{}s, especially to combat against failing GitHub OSS projects.

\section{Background and Related Work}
In this section, we present the background for existing work related to newcomers.
The section is divided into two parts: work related to the sustaining software projects and the barriers that a OSS newcomer faces.

\subsection{Sustaining OSS at Project and Ecosystem level}
Software project sustainability issues have been widely emphasized.
Many software projects have failed frequently due to the insufficient volunteer participation~\cite{Coelho2017FSE}. 
As reported by Valiev et al.~\cite{Valiev2018FSE},  projects should carefully consider the impact that inconsistent participation can have on the ecosystem. Samoladas et al.~\cite{Ioannis2010IST} applied the survival analysis techniques to predict the survivability of the projects. 
Their results quantified the benefit of adding more committers to FLOSS projects.
The findings of Zhou  and  Mockus~\cite{Zhou2015TSE} suggest the importance of initial behaviour and experiences of new participants and outline empirically-based approaches to help the communities with the recruitment of contributors for long-term participation. 
Fang and Neufeld~\cite{Fang2009JMIS} suggest that project success may demand a small number of core developers. 

Projects need mediums or strategies to attract participants to make direct (software coding) contributions apart from indirect (advisory) contributions.
 Qiu et al.~\cite{Qiu2019ICSE} confirmed that social capital is beneficial for prolonged engagement for both genders through the empirical study on Github projects.
On the other hand, we cannot ignore the importance of understanding the role of tools in effective management of the broad network of potential contributors which may lead to code contribution that is more satisfying \cite{M.Zhu2016FSE}. In a recent study, Miller et al.~\cite{Miller2019OSS} found that contributors tend to disengage for different reasons such as the popularity of the project.
In addition, Subramanian~\cite{vikram2020ICSE} found a large fraction of the first pull requests made to an OSS project by developers do not involve writing code, and that contributions are a mixture of trivial and non-trivial changes.

\subsection{Barriers for Newcomers}
Newcomers are important to the survival, long-term success, and continuity of OSS projects~\cite{Kula2019book}. However, newcomers face many difficulties when making their first contribution to a project. OSS project newcomers are usually expected to learn about the project on their own~\cite{Scacchi2002IEE}. Conversely, newcomers to a project, send contributions which are not incorporated into the source code and give up trying~\cite{Steinmacher2015ICSS, Igor2015CSCW}. As discussed by Zhou  and  Mockus~\cite{zhou2010}, the transfer of entire projects to renewal of core developers, participation in OSS projects, present similar challenges of rapidly increasing newcomer competence in software projects. 

Several research activities addressed for reducing the barriers for newcomers previously. Steinmacher et al.~\cite{Steinmacher2014} proposed a developer joining model that represents the stages that are common to and the forces that are influential to newcomers being drawn or pushed away from a project. 
Steinmacher et al.~\cite{Steinmacher2016ICSE} created a portal called FLOSScoach based on a conceptual model of barriers to support newcomers. The evaluation shows that FLOSScoach played an important role in guiding newcomers and in lowering barriers related to the orientation and contribution process. Different from previous research, we conduct a deeper investigation to understand the behaviors of \newcomer~with social interaction, contribution purposes, and target projects.

\label{sec:datasource}
\begin{table*}[ht]
\centering
  \caption{Preliminary Survey Questions}
  \label{tab:data_col1}
  \begin{tabular}{lp{4.5cm}}
    \toprule
    Survey Questions for \newcomer\\ 
    \midrule
    \textbf{Q1) What are their motivation to make a contribution to GitHub?} a. Learning to Code. \\ b. Assignment or Experiment Project. c. Intend to contribute to an Open Source. \\d. Use to showcase my programming skills. e. Others.\\
    \textbf{Q2) Did they have any prior experience contributing to an OSS before GitHub?} (Yes/No)\\
    \textbf{Q3) List their programming knowledge/interests?}\\
  \bottomrule    
\end{tabular}
\end{table*}

\begin{figure*}[t]
    \begin{footnotesize}
    \begin{tikzpicture}
    \begin{axis}[
        align=left,
        y=0.5cm,
        x=0.1cm,
        enlarge y limits={abs=0.25cm},
        symbolic y coords={e. Others,d. Use to showcase my programming skills,c. Intend to contribute to an Open Source,b. Assignment or Experiment Project,a. Learning to Code},
        xlabel={Percentage of Respondents},
        axis line style={opacity=0},
        major tick style={draw=none},
        ytick=data,
        xmin = 0,
        xmax = 100,
        nodes near coords={\pgfmathprintnumber\pgfplotspointmeta\%},
        nodes near coords align={horizontal},
        point meta=rawx
    ]
    \addplot[xbar,fill=gray,draw=none] coordinates {
        (58.9,a. Learning to Code)
        (21,b. Assignment or Experiment Project)
        (82.1,c. Intend to contribute to an Open Source)
        (42.4,d. Use to showcase my programming skills)
        (4.4,e. Others)
        };
    \end{axis}
    \end{tikzpicture}
    \end{footnotesize}
    \caption{Answer to Survey Q1) What are their motivations to make a contribution to GitHub?}
    \label{fig:motivation_to_contribute}
\end{figure*}
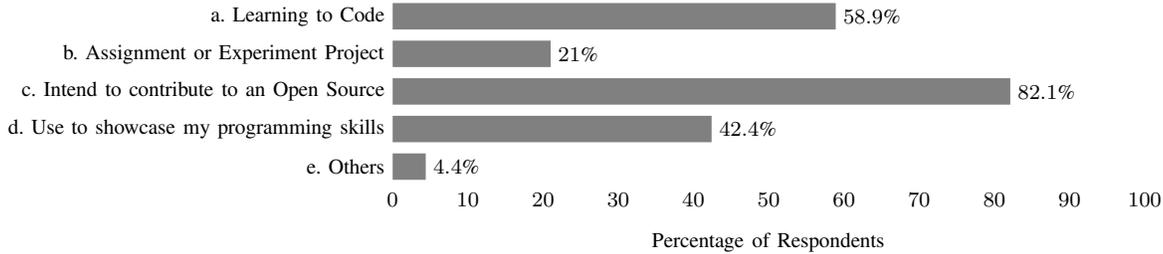

\section{Study Protocols}\label{AA}
In this section, we present the design of our study. 
This section includes the preparation of research questions with their motivation and the description of our data source.

\subsection{Research Questions}
We formulate four research questions that guide our study:

\begin{itemize}
\item \textbf{RQ1: \textit{To what extent does a \newcomer~practice social coding?}}  
Scacchi~\cite{Scacchi2002IEE} showed that newcomers are more likely to learn on their own. 
Our motivation of the first research question is to understand whether or not a \newcomer~tends to collaborate with other users on GitHub.
Based on RQ1, we raise the following hypothesis to confirm the prior work: 
\begin{quote}
  \textit{H1: A \newcomer~is more likely to practice social coding to GitHub.}  
\end{quote}
\item\textbf{RQ2: \textit{What are the kinds of initial contributions that come from a \newcomer{}?}} 
We would like to investigate the typical activities that a \newcomer~is more likely to engage in. Answering this research question will allow us to understand whether a \newcomer~immediately engages into OSS activities, or will first use GitHub for other uses. Our assumption is:
\begin{quote}
    \textit{H2: A contribution to Github repository for a \newcomer~is more likely to add new content.}
\end{quote}
\item\textbf{RQ3: \textit{What kinds of repositories does a \newcomer~target?}}
Kalliamvakou  et  al.~\cite{kalliamvakou2014MSR} showed that most repositories on GitHub are non-software related and are for personal use.
Thus, the motivation is to understand the kinds of projects that attract interest of a \newcomer.
Our assumption is:
\begin{quote}
    \textit{ H3: A \newcomer~is more likely to target software repositories.}
\end{quote}
\item \textbf{RQ4: \textit{What proportion of \newcomer{}s eventually onboard to an OSS project?}} 
In this exploratory research question, we investigate the proportion of \newcomer{}s that eventually onboard to an OSS project.
Answering RQ4 can allow us to understand whether the onboarded \newcomer{}s share similar kinds of initial contributions (RQ2) and repositories (RQ3).
\end{itemize}

\begin{figure}[t]
    \centering
    \begin{footnotesize}
    \begin{tikzpicture}
    \begin{axis}[
        align=left,
        y=0.5cm,
        x=0.07cm,
        enlarge y limits={abs=0.25cm},
        symbolic y coords={Yes,No},
        xlabel={Percentage of Respondents},
        axis line style={opacity=0},
        major tick style={draw=none},
        ytick=data,
        xmin = 0,
        xmax = 100,
        nodes near coords={\pgfmathprintnumber\pgfplotspointmeta\%},
        nodes near coords align={horizontal},
        point meta=rawx
    ]
    \addplot[xbar,fill=gray,draw=none] coordinates {
        (14.5,Yes)
        (85.7,No)
        };
    \end{axis}
    \end{tikzpicture}
    \end{footnotesize}
    \caption{Answer to Survey Q2) Did they have any prior experience contributing to an OSS before GitHub?}
    \label{fig:prior_exp}
\end{figure}
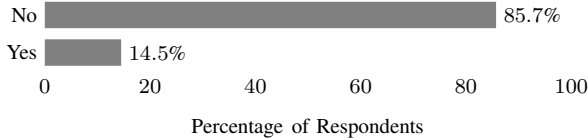

\subsection{Collected Dataset}
Our pilot dataset comes from two different sources, which is a survey questionnaire and a GitHub user dataset.
We now discuss each source and the extraction method to build each dataset. The final dataset will be made publicly available upon the completion of the study.

\subsubsection{Preliminary Survey}
We used the first-contributions project\footnote{\url{https://github.com/firstcontributions/first-contributions}} and its community in GitHub as our data source for collecting \newcomer{}s. Out of the 10,000 \newcomer{}s that we contacted, 208 \newcomer{}s agreed to let us use their GitHub information and verify that they indeed have not contributed to any OSS projects in the past.

Table~\ref{tab:data_col1} details the questions asked for the verification of \newcomer~status.  
To verify our users and ensure that they are indeed first-time contributors to OSS, we created and sent out a questionnaire of Google survey form to the \newcomer{}s. This survey was sent to users through their websites and slack channel\footnote{\url{https://firstcontributors.slack.com/}}. To ensure the privacy of these users, we provide anonymity of the dataset. Our questionnaire is available here at {\url{https://tinyurl.com/r7acxvn}}.

Figure~\ref{fig:motivation_to_contribute} shows the first question (see Table~\ref{tab:data_col1}) response in survey, i.e., 58.9\% of \newcomer motivation for contribution to GitHub is code learning, 21\% is assignments/experiment projects, 82.1\% \newcomer motivation intends to contribute to an OSS project which is the highest, last 42.2\% is to showcase of programming skills. Rest of various motivations for contributions go for others which are 4.4\%. Figure~\ref{fig:prior_exp} details the second question response which shows that up to 85.7\% of \newcomer{}s have no prior experience contributing to an OSS project before GitHub while 14.5\% have prior experience. 
As per response of third question, \newcomer{}s possess the knowledge or interests of various kinds of programming languages.

\subsubsection{GitHub User dataset}
In the second dataset, we match the collected users from the preliminary survey to their GitHub contributions. For this, we mined the GitHub REST API\footnote{https://developer.github.com/v3/}. In the end, we were able to successfully match 208 \newcomer{}s with their 2,985 unique projects.

\section{Execution Plan}
In this section, we present the execution plan of our experiment. We plan to use a mixed approach consisting of both quantitative and qualitative methods.

\subsection{Research Method for RQ1:}
For RQ1, we will use a quantitative method to identify whether or not a first-contribution is social or not.
For this, we use the GitHub user dataset that includes all the first contributions of 208 candidates.
Our strategy to identify each contribution as either social or not social by checking if the project has commits from other developers. We plan to use the command \texttt{git-blame}, which shows the code revision and  each line of a file that author last modified\footnote{\url{https://git-scm.com/docs/git-blame}}.

\subsection{Research Method for RQ2:}
For RQ2, we will use a qualitative method to identify the different kinds of a first-contribution. We will manually classify the first contributions across the 208 \newcomer s using the coding guide proposed by Hattori and Lanza~\cite{Hattori2008ASE}.

Similar to Hata et al.~\cite{Hata2019ICSE}, the agreement of the coding guide will be performed using a kappa agreement \cite{viera2005family}. Kappa result is interpreted as follows: values $\leq$ 0 as indicating no agreement and 0.01–0.20 as none to slight, 0.21–0.40 as fair, 0.41– 0.60 as moderate, 0.61–0.80 as substantial, and 0.81–1.00 as almost perfect agreement.

\subsection{Research Method for RQ3:}
For RQ3, we will use a qualitative method similar to RQ2, to identify the different kinds of repositories that receive contributions from \newcomer s. 
For a more in-depth analysis of the repositories, we distinguish between fork and non-forked.
Our assumption is that repositories follow the typical pull-based workflow, which is first a fork and then submit a pull-request to that original repository.
We will then manually classify these repositories based on the coding guide proposed by Borges et al.~\cite{Borges2016ICSME} and  Kalliamvakou  et  al.~\cite{kalliamvakou2014MSR}.
Output will be a coding guide. Our sample will consist of 95 percent confidence level and 0.05 confidence interval value. Similar to RQ2, the agreement will be performed using a kappa agreement.
\subsection{Research Method for RQ4:}
For RQ4, our approach includes two steps. 
First, we use the GitHub user dataset to match all unique projects of 208 \newcomer{}s with curated dataset of engineered software projects provided by Munaiah  et  al.~\cite{meiESE2016}.
For each \newcomer, we match their projects against the curated dataset.
We regard a \newcomer~as onboarded as if at least one of their projects is indeed one of those engineered software projects.
For a qualitative analysis, we will validate newcomer barriers \cite{igor2014IST} by conducting a user feedback survey.

\section{Analysis Plan}
In this section, we present the analysis of our results. 
For all research questions, we now present the basis of the coding guides used in the study.

\subsection{Analysis for RQ1:}
The results of RQ1 will be presented using a table of descriptive statistics and a pivot graph.

\subsubsection{Descriptive Statistics}
We will present a pivot chart to show the frequency of social and non-social contributions, x-axis will represents social and not-social contributions and y-axis will represents the frequency count of contributions. 

\subsubsection{Testing significance}
For significance, we will test our hypothesis \textit{ (H1): A \newcomer~is more likely to practice social coding to GitHub.} Based on the results, we apply the statistical test for normality test, e.g., using the Shapiro-Wilk procedure when using non-parametric tests with Cliff's delta statistics. The effect size is considered: (1) negligible if d $<$ 0.147, (2) small if d $<$ 0.33 (3) medium if d $<$ 0.474 (4) otherwise large.

\subsection{Analysis for RQ2:}
\subsubsection{Manual Classification} We will follow the coding scheme proposed by Hattori and Lanza~\cite{Hattori2008ASE} for guiding the classification of contribution purposes:
\begin{itemize}
    \item{\underline{Forward Engineering:}} development activities are those related to incorporation of new features and implementation of new requirements.
    \item{\underline{Re-engineering:}} maintenance activities are related to refactoring, redesign and other actions to enhance the quality of the code without properly adding new features
    \item{\underline{Corrective Engineering:}} maintenance activities
    handle defects, errors and bug in the software.
    \item{\underline{Management:}} maintenance activities are those unrelated to codification, such as formatting code, cleaning up, and updating documentation.
\end{itemize}

To show the results, we will draw a histogram plot which reflects the diversity of different contribution purposes. The plot will clearly show what kinds of motivation is the most frequent for \newcomer{}s.

\subsubsection{Testing significance}
To test our hypothesis \textit{(H2): A contribution to Github repository for a \newcomer~is more likely to add new content}, similar to the analysis for RQ1, we apply the Shapiro-Wilk procedure to verify the significant difference and use the Cliff's delta statistics to measure the non parametric effect size.

\subsection{Analysis for RQ3:}
\subsubsection{Manual Classification} For RQ3, our coding scheme has two labelled categories.
The first is the kinds of software domain based on Borges et al.~\cite{Borges2016ICSME}:
\begin{itemize}
    \item{\underline{Application Software:}} systems that provide functionalities to end-users, like browsers and text editors.
    \item{\underline{System Software:}} systems that provide services and infrastructure to other systems, like operating systems, middleware, servers, and databases.
    \item{\underline{Web-based-application, libraries, and frameworks.}}
    \item{\underline{Non-web libraries and frameworks.}}
    \item{\underline{Software tools:}} systems that support software development tasks, like IDEs, package managers, and compilers.
    \item{\underline{Documentation:}} repositories with documentation, tutorials, source code examples, etc.
\end{itemize}
The second is the types of non-software categories based on Kalliamvakou et al.~\cite{kalliamvakou2014MSR}:
\begin{itemize}
    \item{\underline{Experimental:}} repositories includes demos, samples, test code and tutorial examples.
    \item{\underline{Storage:}} category includes repositories documents and files for personal use, such as presentation slides, resumes, e-books, music files etc.
    \item{\underline{Academic:}} class and university research projects come under this category.
    \item{\underline{Web:}} under this category we classify websites and blogs.
    \item{\underline{No longer accessible:}} repositories that gave 404 error will classify under this category.
    \item{\underline{Empty:}} repositories containing only a license file, a gitignore file, a README file, or no files at all were place under this category.
\end{itemize}

We draw two histograms separately for fork and non-fork projects to show the distribution of projects that receive contributions from \newcomer s.

\subsubsection{Testing significance}
To test our hypothesis \textit{(H3): A \newcomer~is more likely to target software repositories.}, similar to the analysis for RQ1, we apply the Shapiro-Wilk procedure to verify the significant difference and use the Cliff's delta statistics to measure non-parametric effect size.

\subsection{Analysis for RQ4:}
For the result of RQ4, we will use a table of descriptive statistics and a pivot graph similar to RQ1 analysis.
\subsubsection{Descriptive Statistics} 
We plan to conduct two analysis. First, we present the proportion of \newcomer s that will onboard compared to those that do not. Furthermore, we will provide statistics such as the time taken to onboard and the number of projects. Then, we use the classifications of RQ2 and RQ3 to make observations about commonalities of initial contributions and the types of repositories that \newcomer{}s that eventually onboard an OSS project.
In terms of the survey, we will ask participants to indicate their level of agreement on a five-point Likert scale (from "strongly disagree" to "strongly agree") for the following barriers \cite{igor2014IST}:
\begin{itemize}
    \item {\underline{Social Interaction:}} barriers related to the manner in which \newcomer{}s interact with community members.
    \item {\underline{Newcomer Previous Knowledge:}} barriers related to the experience of \newcomer{}s regarding the project and the manner in which they show experience when joining the projects.
    \item {\underline{Finding a Way to Start:}} barriers related to difficulties that \newcomer{}s face when trying to find the right way to start contributing.
    \item {\underline{Technical Hurdles:}} barriers imposed by the project when \newcomer{}s are dealing with the code.
    \item {\underline{Documentation:}} barriers related to documentation problems like outdated documentation, too much documentation, unclear code comments.
\end{itemize}
\subsubsection{Testing significance}
Since RQ4 is exploratory, there is no hypothesis to test.

\section{Implications}
We summarize our implications with the following take-away messages for the key stakeholders:
\begin{enumerate}
    \item \textit{\Newcomer-}
Our research helps \newcomer{}s understand the kinds of contributions they perform before onboarding to real OSS project. 
Conversely, we can reveal barriers on why some newcomers never end up contributing to an OSS projects.
    \item \textit{OSS projects-} 
Findings will reveal insights into what contributions may attract a \newcomer.
OSS projects may get benefit from it, if they offer the right contributions for right \newcomer{}s.
    \item \textit{ Researchers-} Non-software repositories that are personal have always been regarded as perils, we explore these projects to understand what a \newcomer~is trying to advertise in these projects, and to what extent does it lead to onboarding to an OSS project.
\end{enumerate}

\section*{Acknowledgement}
This work is supported by Japanese Society for the Promotion of Science (JSPS) KAKENHI Grant Numbers 18H04094, 18H03221, 18KT0013, and 20K19774.

\bibliographystyle{IEEEtran}
\bibliography{ref}

\begin{thebibliography}{10}
\providecommand{\url}[1]{#1}
\csname url@samestyle\endcsname
\providecommand{\newblock}{\relax}
\providecommand{\bibinfo}[2]{#2}
\providecommand{\BIBentrySTDinterwordspacing}{\spaceskip=0pt\relax}
\providecommand{\BIBentryALTinterwordstretchfactor}{4}
\providecommand{\BIBentryALTinterwordspacing}{\spaceskip=\fontdimen2\font plus
\BIBentryALTinterwordstretchfactor\fontdimen3\font minus
  \fontdimen4\font\relax}
\providecommand{\BIBforeignlanguage}[2]{{%
\expandafter\ifx\csname l@#1\endcsname\relax
\typeout{** WARNING: IEEEtran.bst: No hyphenation pattern has been}%
\typeout{** loaded for the language `#1'. Using the pattern for}%
\typeout{** the default language instead.}%
\else
\language=\csname l@#1\endcsname
\fi
#2}}
\providecommand{\BIBdecl}{\relax}
\BIBdecl

\bibitem{park2009VISSOFT}
Y.~Park and C.~Jensen, ``Beyond pretty pictures: Examining the benefits of code
  visualization for open source newcomers,'' in \emph{VISSOFT}, 2009.

\bibitem{Fang2009JMIS}
Y.~Fang and D.~Neufeld, ``Understanding sustained participation in open source
  software projects,'' \emph{J. Manage. Inf. Syst.}, 2009.

\bibitem{Steinmacher2014}
I.~Steinmacher, M.~A. Gerosa, and D.~Redmiles, ``Attracting, onboarding, and
  retaining newcomer developers in open source software projects,'' in
  \emph{CSCW}, 2014.

\bibitem{Valiev2018FSE}
M.~Valiev, B.~Vasilescu, and J.~Herbsleb, ``Ecosystem-level determinants of
  sustained activity in open-source projects: A case study of the {PyPI}
  ecosystem,'' in \emph{FSE}, 2018.

\bibitem{Coelho2017FSE}
J.~Coelho and M.~T. Valente, ``Why modern open source projects fail,'' in
  \emph{FSE}, 2017.

\bibitem{igor2014IST}
I.~Steinmacher, M.~A. Graciotto~Silva, M.~A. Gerosa, and D.~Redmiles, ``A
  systematic literature review on the barriers faced by newcomers to open
  source software projects,'' \emph{IST}, 2014.

\bibitem{Ioannis2010IST}
I.~Samoladas, L.~Angelis, and I.~Stamelos, ``Survival analysis on the duration
  of open source projects,'' \emph{IST}, 2010.

\bibitem{Zhou2015TSE}
M.~{Zhou} and A.~{Mockus}, ``Who will stay in the floss community? modeling
  participant’s initial behavior,'' \emph{TSE}, 2015.

\bibitem{Qiu2019ICSE}
H.~S. Qiu, A.~Nolte, A.~Brown, A.~Serebrenik, and B.~Vasilescu, ``Going farther
  together: The impact of social capital on sustained participation in open
  source,'' in \emph{ICSE}, 2019.

\bibitem{M.Zhu2016FSE}
J.~Zhu, M.~Zhou, and A.~Mockus, ``Effectiveness of code contribution: From
  patch-based to pull-request-based tools,'' in \emph{FSE}, 2016.

\bibitem{Miller2019OSS}
C.~Miller, D.~G. Widder, C.~K{\"a}stner, and B.~Vasilescu, ``Why do people give
  up flossing? {A} study of contributor disengagement in open source,'' in
  \emph{OSS}, 2019.

\bibitem{vikram2020ICSE}
V.~N. Subramanian, ``An empirical study of the first contributions of
  developers to open source projects on github,'' in \emph{ICSE Companion},
  2020.

\bibitem{Kula2019book}
R.~G. Kula and G.~Robles, \emph{The Life and Death of Software
  Ecosystems}.\hskip 1em plus 0.5em minus 0.4em\relax Springer, 2019.

\bibitem{Scacchi2002IEE}
W.~{Scacchi}, ``Understanding the requirements for developing open source
  software systems,'' \emph{IEE Proc. Soft}, 2002.

\bibitem{Steinmacher2015ICSS}
I.~{Steinmacher}, T.~U. {Conte}, and M.~A. {Gerosa}, ``Understanding and
  supporting the choice of an appropriate task to start with in open source
  software communities,'' in \emph{HICSS}, 2015.

\bibitem{Igor2015CSCW}
I.~Steinmacher, T.~Conte, M.~A. Gerosa, and D.~F. Redmiles, ``Social barriers
  faced by newcomers placing their first contribution in open source software
  projects,'' in \emph{CSCW}, 2015.

\bibitem{zhou2010}
M.~Zhou and A.~Mockus, ``Growth of newcomer competence: Challenges of
  globalization,'' in \emph{FoSER}, 2010.

\bibitem{Steinmacher2016ICSE}
I.~Steinmacher, T.~U. Conte, C.~Treude, and M.~A. Gerosa, ``Overcoming open
  source project entry barriers with a portal for newcomers,'' in \emph{ICSE},
  2016.

\bibitem{kalliamvakou2014MSR}
E.~Kalliamvakou, G.~Gousios, K.~Blincoe, L.~Singer, D.~M. German, and
  D.~Damian, ``The promises and perils of mining {GitHub},'' in \emph{MSR},
  2014.

\bibitem{Hattori2008ASE}
L.~P. Hattori and M.~Lanza, ``On the nature of commits,'' in \emph{ASE}, 2008.

\bibitem{Hata2019ICSE}
H.~Hata, C.~Treude, R.~G. Kula, and T.~Ishio, ``9.6 million links in source
  code comments: Purpose, evolution, and decay,'' in \emph{ICSE}, 2019.

\bibitem{viera2005family}
A.~J. Viera and J.~M. Garrett, ``Understanding interobserver agreement: the
  kappa statistic,'' \emph{Family Medicine}, 2005.

\bibitem{Borges2016ICSME}
H.~{Borges}, A.~{Hora}, and M.~T. {Valente}, ``Understanding the factors that
  impact the popularity of {GitHub} repositories,'' in \emph{ICSME}, 2016.

\bibitem{meiESE2016}
N.~Munaiah, S.~Kroh, C.~Cabrey, and M.~Nagappan, ``Curating github for
  engineered software projects,'' \emph{EMSE}, 2016.

\end{thebibliography}
\end{document}